# Segregation of two seed growth patterns with fractal geometry

Deepak N. Bankar[1], P. M. Gade[2], A.V. Limaye[1] and A. G. Banpurkar[1*]

[1]*Center for Advanced Studies in Materials Science and Solid State Physics,*
*Department of Physics, University of Pune, Pune - 411 007, India*

[2]*Centre for Modeling and Simulation, University of Pune, Pune - 411 007, India*

We study the generalized diffusion-limited aggregates (DLA), with two seeds placed at distance $d$ lattice units and investigate the probability $p(d)$ that the patterns generated from those seeds get connected. In this model, one can vary the parameter α, and get a range of patterns from fractal-DLA to compact one. For a fractal-DLA, $p(d)$ decays rapidly with $d$, the decay is slower for compact pattern in which case $p(d) \approx 1$ for all practical distances. We demonstrate a similar phenomenon experimentally in viscous fingering and electrochemical deposition with two injection points/cathodes.

Aggregation of particles is a common phenomenon in many branches of applied science and is of interest from commercial and fundamental viewpoints [1]. Studies in aggregation and pattern formation have been greatly advanced due to novel experiments as well as the rapid development of simulation models. Among several models, diffusion-limited aggregation (DLA) model developed by Witten and Sander [2] has arguably been the most important one. The morphology of the DLA patterns is very much similar to those obtained in many natural systems as well as experiments such as bacterial growth, solidification, viscous-fingering etc [1, 3].

In DLA simulation and the experiments, the growth is presumed to begin from a *single* immobile seed, which is not always true. It may start simultaneously from several adjacent seeds/sites and resulting patterns could be completely different. In this work, we study in detail, the connectivity of such patterns as a function of their morphology and inter-seed separation. We believe that apart from theoretical interest, these studies could have several practical applications ranging from ecology to the formation of extended network. For instance, one would see a dramatic change in the conductivity of the structure, whose neighboring clusters of conducting particle do connect. Witten and Meakin, to some extent, have attempted a similar issue in their work on DLA for multiple growth sites grown on a square lattice. They observed that DLA clusters simultaneously grown on two seeds separated by nine vacant lattice-sites remain unconnected *during their simulation time* [4]. This issue also finds a passing mention in the subsequent study on a model for biological pattern formation by P Meakin [5]. In this model, randomly chosen active site gets occupied with probability $P \sim c^\varepsilon$, where $c$ is the local concentration of nutrient diffusing from a surrounding exterior source and consumed by the growing cluster and $\varepsilon$ is some parameter. Simulations on square lattice with two seeds separated by ten lattice-sites showed that the clusters remained disjoint for $\varepsilon = 1.0$ and $2.0$ while they connected for $\varepsilon = 0.5$.

Experimentally, H Fujikawa *et al.* inoculated *Bacillus Subtilis* bacterial strains simultaneously at two closely spaced points on the surface of an agar plate. Their colonies did not join under moderate nutrient concentration but fused together for high nutrient concentration [6]. We recall that under starvation conditions, bacterial colonies show growth similar to a DLA model [6,7]. In ecology also, a similar problem has been studied and researchers have put forth a Competitive Exclusion Principle (CEP). CEP is essentially related to the ability of competing species to coexist in a given niche. May [8] defines it, as the instance where two species which makes their living in individual way being unable to exist in a stable fashion. Our work identifies a common theme explaining known observations and our studies. This theme could be viewed as the generalization of CEP.

In case of patterns grown from two separate seeds, naive expectation would be that the patterns would connect together if one waits long enough. *Systematic studies of this problem are required for checking if this is really true.* In this letter, we systematically study the behavior of probability $p(d)$ that the patterns generated from two seeds get connected for a range of inter-seed distances $d$ using generalized DLA model. We also experimentally demonstrate a similar phenomenon in viscous fingering and electrochemical deposition with two injection-points/cathodes. All these studies seem to have interesting common underlying features, which could be described as generalization of CEP.

While inter-seed distance obviously affects connectivity of cluster, we find that fractal dimension $D_F$ of clusters is also crucial. A DLA model can be generalized by introducing a parameter called sticking probability [9,10]. It allows us to vary $D_F$ of clusters. In the generalized DLA model, particles stick to the cluster on visiting active site with sticking probability $s$, $s = \alpha^{3-B}$, where $\alpha$ is some positive parameter ($0 < \alpha \leq 1$) and $B$ is the number of nearest-neighbor occupied sites in the cluster [9]. For $\alpha = 1$ the model generates DLA patterns which are fractal. For smaller values of α (α→0), it generates compact morphology with $D_F \sim 2$ since active sites with $B = 3$ are more likely to get occupied than those with $B = 1$ [10].

We simulated generalized DLA model on a 2-D square lattice with two seeds separated by a $d$ lattice units. They are placed at the lattice points ($[-d/2]$, 0) and ($[d/2]$, 0), where $[x]$ stands for largest integer $\leq x$. The effects of asymmetric placement of seeds (in case of odd $d$ values) relative to the launching circle were expected to be insignificant since initial launching circle radius $r = 100$ lattice units was much larger than $d$. (We carefully increase $r$ as the cluster grows.) A particle launched from the lattice-site nearest to the randomly chosen point on the circumference of launching circle carries a random walk and becomes a part of the cluster with sticking probability $s = \alpha^{3-B}$. If a walker becomes a part of both the clusters, the clusters are said to be connected. We vary α from 0.1 to 1 and $d$ from 2 to 9 lattice units.

First we discuss the representative morphology of generalized DLA structures generated from two seeds. In Fig. 1(a-c) we show the patterns for α =1.0, 0.7 and 0.3 respectively at inter-seed distance $d = 9$ lattice units. The fractal dimension of the corresponding patterns are $D_F \sim 1.67$, 1.74 and 1.84. This shows that as α decreases $D_F$ increases, which are consistent with the result by Banavar *et al*. [10]. In all three cases, the patterns from two seeds are clearly not connected. This segregation is primarily due to the well-known shielding effect [4], which prevents a walker from penetrating in the empty spaces between the clusters. However, as α→ 0, a walker can penetrate into deeper voids results into a compact morphology and hence in a higher probability of connectivity.

One can argue that the patterns, which do not seem to meet at certain stage, could eventually meet after some additional particles are accumulated in the cluster. Obviously, no simulation can run for infinite time and systematic approach must be employed to find the asymptotic probability. We define a quantity $P(d, N)$ as the probability of connectivity between two clusters where, $N$ is the total number of particles in the clusters. For each set of α and $d$ values, we generate 1000 configurations. In each configuration, clusters were observed for connectivity till $N_{max}$ particles are added. $P(d, N)$ is then the fraction of 1000 configurations which get connected on or before the accumulation of $N$ particles in the clusters. Figure 2 shows a plot of $P(d, N)$ as a function of $N$ on a semi-logarithmic scale for some α and $d$ values. As expected, for larger $d$, we do not see any connectivity till adequate particles are aggregated. It is evident that, as $N \to \infty$, the probability $P(d, N)$ saturates in all the cases. If the saturation value is not reached for $N_{max}$ particles, we try a higher value of $N_{max}$ till the saturation is clearly obtained. We denote this asymptotic probability by $p(d)$. As expected, for larger *d*, the saturation occurs at higher value of *N* and saturation value is smaller. The plot also shows that *p(d)* increases with decreasing α as expected from the reduction in shielding effect.

In Fig. 3, we plotted the asymptotic probability, *p(d)* against the *d* for different α values. The plot shows that there is a rapid decay of *p(d)* with *d* for α values close to 1. However the decay is slower with decreasing α. We fitted this data points using the Gaussian form, $p(d) = \frac{1}{\sqrt{2\pi}\,\xi} \exp\left(\frac{-d^2}{2\xi^2}\right)$. The continuous line show the fitted curve to the data points and it is clear that the fit works very well. It is also seen that the standard deviation, ξ increases with decreasing α. For very small α, when cluster is compact, ξ is very large. We conjecture that for compact structures, $\xi \to \infty$ and $p(d) \approx 1$ for all values of *d*. Thus there is a significant difference between connectivity properties of fractal DLA and compact structures.

The Gaussian nature of the *p(d)* may be primarily due to that the probability of two clusters getting connected is essentially related to overlap integral of the probability distribution of the two clusters. It is known that the radial growth probability of the individual clusters show Gaussian behavior [11] and for Gaussians, overlap integral is again a Gaussian.

We further tried to show that for compact cluster $p(d) \approx 1$ for all distances using another growth model like Ballistic driven aggregation (BDA) with 4-sided rain [12]. Certainly this model leads to a compact cluster. The model with two seeds is as follows: the two seeds separated by distance *d* are placed at lattice points on the square lattice of size $N \times N$. A particle starts at any site on the perimeter of the square and follows a linear path. If perimeter site is on right (left) edge, particle moves to left (right). Similarly if the chosen site is on top (bottom) of the square, it moves to bottom (top). It gets attached to the cluster on lattice whenever it visits any active site. We observe that the clusters connect with probability one for any distance *d*, giving *p(d)* =1 for all distances.

These observations also raise interesting question. *Does this merging phenomenon have some critical dimension above which the clusters created from different seeds certainly merge?* Since we observe that compact like 2-*D* clusters do merge with probability 1, would the growing clusters certainly meet each other when underlying dimension is greater than 2? To settle this question, we carried out simulations for a DLA in 3-*D* with two seeds placed at distance *d* = 2 lattice unit. We observe that though the time required for convergence to asymptotic value is much longer, the asymptotic value is certainly less than one (see Fig. 4). This implies that the *crucial factor* that distinguishes between the patterns that (certainly) connects and those remain segregated is whether the underlying clusters are fractal or compact in a given embedding Euclidean dimension $D_E >1$. While compact like structures in all embedding Euclidean dimension $D_E > 1$ are likely to merge and the fractal structures do not merge.

This phenomenon can be illustrated experimentally as well. As mentioned before, patterns generated in several experimental situations closely resemble DLA. One such case is viscous fingering in a radial Hele-Shaw cell for high flow rate [13]. When a low viscosity fluid is injected into a high viscosity fluid, it advances in the form of fingers. This phenomenon is called as viscous fingering [14] and is of relevance to problems like indirect recovery of oil. Thus our studies hence could have practical implications. Hele-Shaw cell consists of two horizontal glass plates with uniform spacing between them. High viscosity fluid is filled in this gap and low viscosity fluid is injected through a hole drilled into a top glass plate. In the present study, glycerine ($\mu = 850$ cP) was used as high viscosity fluid and air was used as low viscosity fluid. Smooth plexi-glass plates of dimension $60 \times 60$ cm$^2$ and thickness of 10 mm were used to construct the cell. Uniform plate separation of 0.4 mm was maintained using Teflon spacers. Two holes, each of diameter 2 mm were drilled into the top plate at a distance *d*. Air, is *simultaneously* injected through these holes at constant pressure. Viscous fingering patterns were realized for various *d* values (*d* = 2 to 7 cm) and at different air pressures. Fingering events were video-recorded using CCD camera (pixels resolution 762×582). Images were digitized and analyzed using public domain UTHAS Image Tool software.

Figure 5 shows the morphology of the viscous fingering patterns plotted for different areal-flow rate and distance *d*. It is clearly seen that there exist two distinct regimes in the fingering patterns. For low flow rate and small *d*, the two patterns get connected. However, for higher flow rates and larger *d* values, patterns remain segregated. The fractal dimension $D_F$ of the connected viscous fingering patterns was found to be close to 2, while those for the segregated pattern were found to be close to 1.74. This one again supports our inference that the two seed patterns merge when they are compact and remain segregated when they are fractal. Another interesting observation is that for the segregated viscous fingering patterns the average separation between two clusters is found to be *d*/2. This feature was also seen in the simulation of DLA clusters remaining unconnected.

Electrodeposition is another physical process, which closely resembles DLA [15]. We study the Zn electro-deposits in Hele-Shaw geometry for two-cathodes positioned at a distance of 0.5 mm and both are connected to a negative potential of 5.6 *V* with respected to ring anode. The two patterns remain segregated for moderate to low electrolyte concentration and merge at high electrolytic concentration (See Fig. 6). We can see that the patterns, which do not merge show fractal geometry while the other one is more compact. We would like to point out that, to the best of our knowledge, these are the first experimental studies on viscous fingering with two-point injection or electrochemical deposition with two cathodes.

These results can be interpreted from a broader perspective. We argue that these are the cases of segregation induced by competition for limited resource. When we have a fractal growth, percentage of occupied sites is decreases with increase in cluster radius. In fact, it is zero asymptotically. Thus the resource available for growth is limited and is reflected in segregation. For a Hele-Shaw cell, the finger width of advancing tip reduces with increasing fingertip velocity [16]. Thus resource available for growing tip reduces. The Electro-deposits remain segregated for low concentration, *i.e.* for a limited availability of the resource. We can understand previous experimental and theoretical results in the same spirit. H Fujikawa *et al.* showed that under starvation condition, which is a case of limited resource, the bacterial colonies show fractal growth and remain segregated [6]. Meakin's model in which probability of cluster growth is proportional to $c^\varepsilon$ ($c < 1$), the patterns connect only for $\varepsilon = 0.5$ and not for $\varepsilon = 1$ and 2. Thus segregation occurs at weak growth probability. The segregated clusters are fractal while the connected cluster displays a compact morphology [5].

Growth models have been proposed for phenomena ranging from tumor growth to forest fires. It is certainly an important question to know if the adjacent tumors merge and grow together or if the forest fire spreads. Thus this study could have important practical implications. However, on the quantitative side, there are questions, which merit further investigation. The Gaussian dependence of probability of connectivity as a function of distance needs to be studied further. We have also observed that mean distance of separation between two clusters is *d/2* in simulations as well as in experiments on viscous fingering. Even these quantitative features could be generic and detailed studies are needed to establish it.

To conclude, we have studied the connectivity properties of several growth phenomena from theoretical and experimental perspective. We observe that the compact patterns connect together, while the fractal patterns do not. Our work can explain previous studies and puts them in a perspective. All these studies seem to point towards a generalized competitive exclusion principle (CEP) for pattern growth. We conjecture that the competition for limited resources (which also results in fractal growth) induces segregation in growth processes initiated from multiple seeds.

One of the authors (AGB) would like to thank DST (Govt. of India) for the financial support under the grant: **SR/FTP/PS-43/2001.**

*Electronic address: agb@physics.unipune.ernet.in

**Figure captions:**

FIG. 1: Representative morphologies of the DLA patterns generated using two seed configuration for $d = 9$ lattice units. (a) $\alpha = 1.0$ and fractal dimension $D_F = 1.67$ (b) $\alpha = 0.7$; $D_F = 1.74$ and (c) $\alpha = 0.3$; $D_F = 1.84$.

FIG. 2: The probability of connectivity $P(d, N)$ versus number of particles $N$ plotted for $d = 2$ ($\alpha = 1.0$ [O]; 0.7 [◇]; 0.4 [△]; 0.1 [▽]) and $d = 9$ ($\alpha = 0.3$ [····]; 0.2 [----]; 0.1 [—]).

FIG. 3: Asymptotic probability of connectivity $p(d)$ is plotted against inter-seed distance $d$ for various $\alpha$ values.

FIG. 4: Plot shows the variation of $P(2, N)$, against $N$ for 3-$D$ clusters generated using two-seed configuration on cubic lattice.

FIG. 5: Connectivity phase diagram for two-point injection viscous fingering in radial Hele-Shaw cell for air-glycerine system. Line is drawn to guide the eye for separation between two regimes.

FIG. 6: Zn electrodeposits using aqueous $ZnSO_4$ solution for two-cathodes. Connected and segregated patterns observed for electrolyte concentrations (a) 0.02 M (b) 0.01 M, respectively. Scale is 3 mm in length.

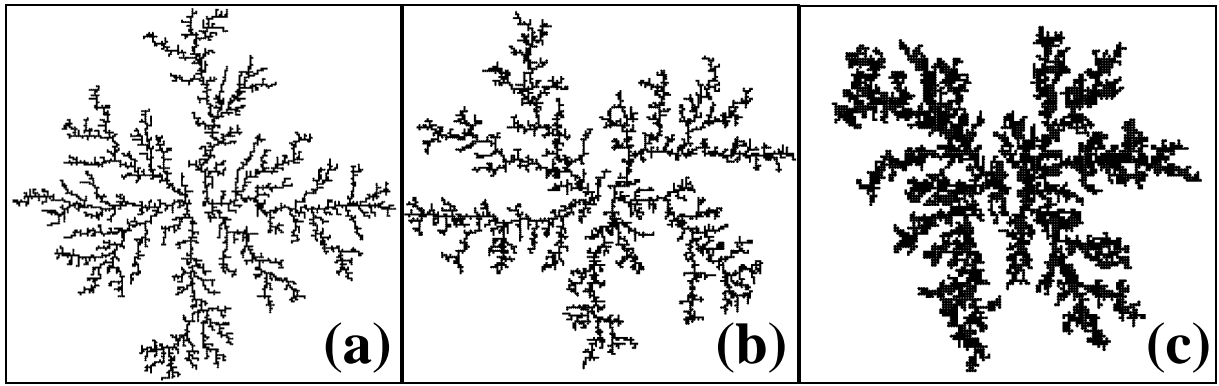

FIG. 1

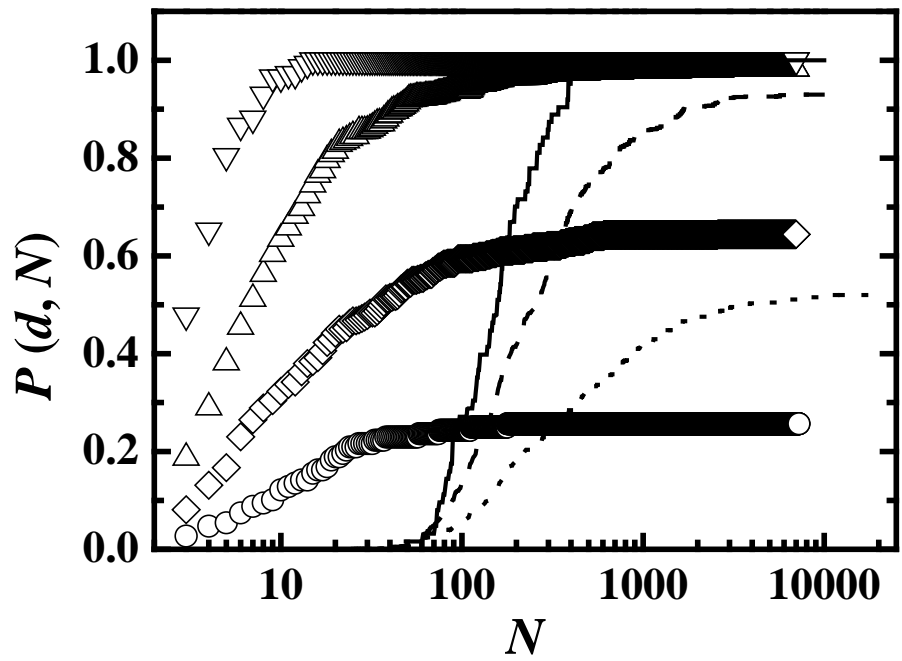

FIG. 2

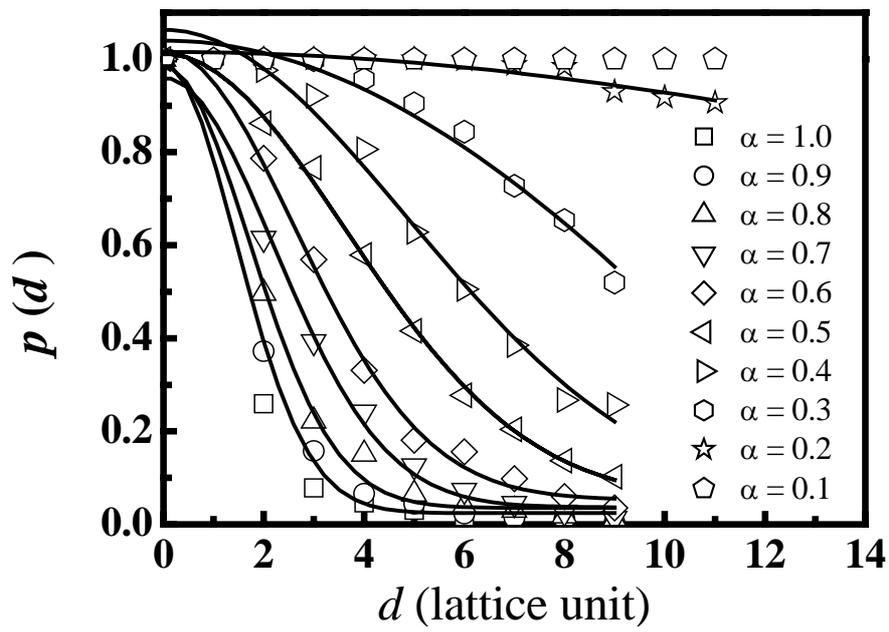

FIG. 3

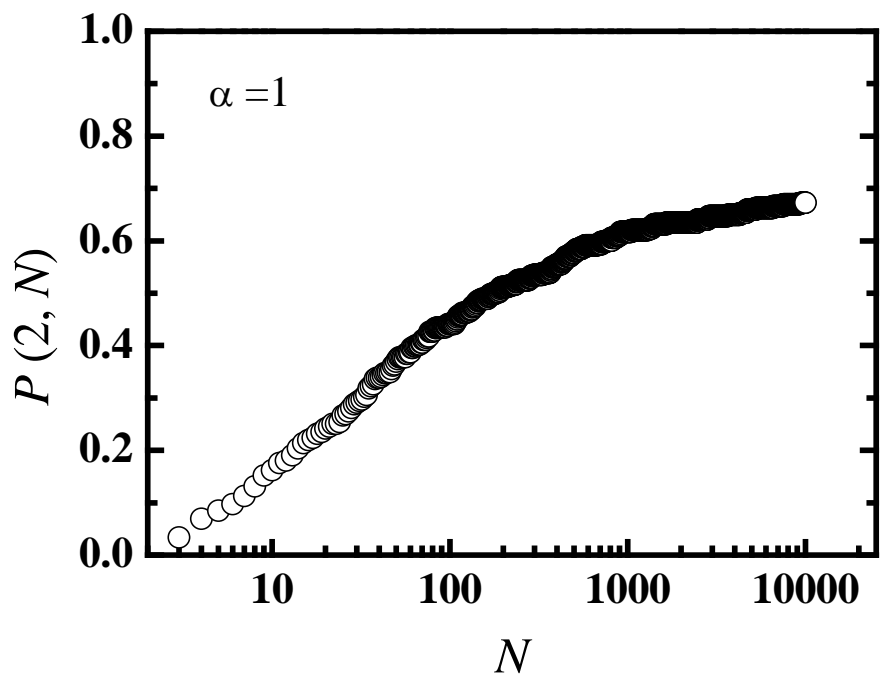

FIG. 4

Segregation of two seed growth patterns with fractal geometry      Deepak N. Bankar et al.

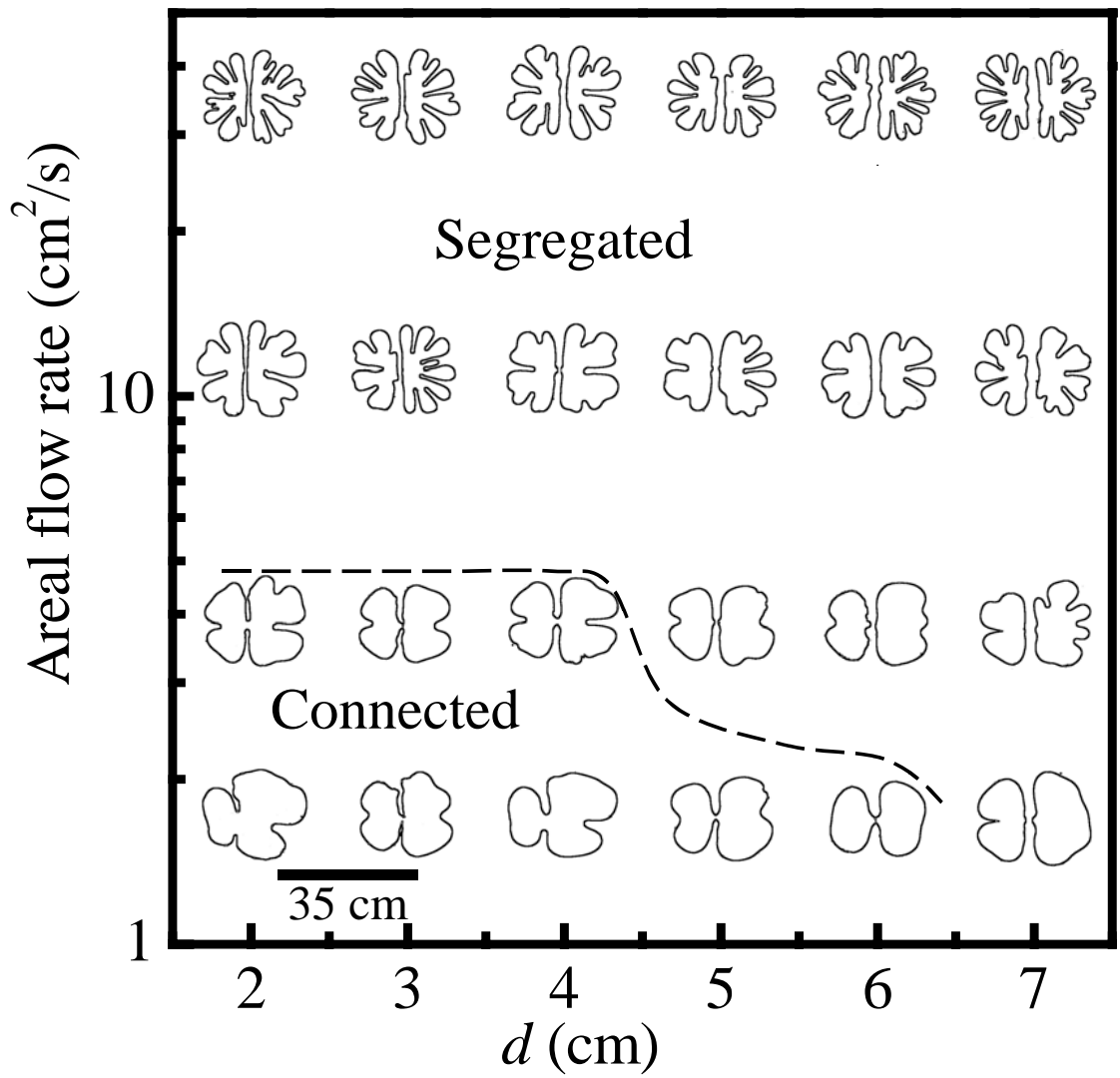

FIG. 5

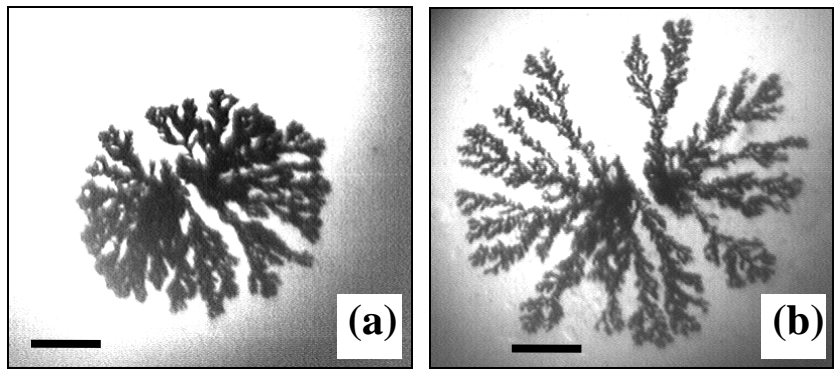

FIG. 6